\documentclass[11pt]{article}             
\title{State-extended uncertainly relations and tomographic inequalities as quantum system state characteristics}
\author{V.~N.~Chernega, V.~I.~Man'ko
        \\$^{\dag}$ P.N.~Lebedev Physical Institute, Russian Academy of Sciences\\
        Leninskii Prospect, 53, Moscow 119991, Russia \\
        \\Emails: vchernega@gmail.com, manko@sci.lebedev.ru}

\begin{document}

\maketitle

\begin{abstract}
Some inequalities for probability vector are discussed. The
probability representation of quantum mechanics where the states are
mapped onto probability vectors (either finite or infinite
dimensional) called the state tomograms is used. Examples of
inequalities for qudit tomograms and a state extended uncertainly
relation are considered. Tomographic cumulant related to photon
state tomographic probability distributions is introduced and it is
used as parameter of the state nongaussianity.
\end{abstract}

\section{Introduction}
Recently ~\cite{CherManko5, CherManko6, PS} the tomographic probability representation of
quantum states ~\cite{Mancini96,IbortPS} was used to study some
uncertainly relations introduced in ~\cite{Trif,Trif2} so called state extended
uncertainly relations. These relations were presented in the form of
integral inequalities ~\cite{Bellini1} for measurable optical tomograms
~\cite{SolimenoPorzio, Raymer, Bellini2}. The aim of our work is to consider some
other inequalities which can be obtained for any probability vectors
and to apply the inequalities to tomographic probability distributions
describing the quantum states of photons and qudits. We consider
recently found new uncertainly relations for arbitrary observables ~\cite{Trif,Trif2} and present the
relations in the form of inequalities for measurable optical photon
tomograms. Also the simple inequalities available for probability
vectors we use to get the inequalities for spin tomographic
probability distributions. Another aim is to introduce the cumulant
related to optical tomogram as a characteristics of the photon state
gaussianity. The paper is organized as follows. In next section 2 we present the state extended uncertainty relations in the form of integral inequalities for the photon state tomograms. In section 3 we remind some properties of probability vectors and discuss the linear maps of such vectors.  The inequalities for Shanon entropy associated with a probability distribution considered as probability vector are studied in section 4. New entropic inequalities are obtained for qudit tomograms in section 5. The tomographic cumulant and corresponding integral inequality expressed in terms of optical tomogram of quantum state is suggested to be used as nongaussianity parameter in experiments on homodyne photon detection in section 6. Conclusion and prospectives are presented in section 7.

\section{State-extended uncertainly relations}

In our previous work~\cite{PS},  some of the
state-extended uncertainly relations ~\cite{Trif}
were presented in tomographic form suitable for
experimental check. The state-extended position and momentum
uncertainly relations were confirmed in experiments
with homodyne photon detection in \cite{Bellini1}.
Now we consider the other state-extended uncertainly relations;
namely, we study the inequality
\begin{eqnarray}\label{eq.dd1.39}
\left((\Delta A(\psi_1))^2+\langle \psi_1|A|\psi_1\rangle^2\right)
\left((\Delta
A(\psi_2))^2+\langle \psi_2|A|\psi_2\rangle^2\right)\geq|\langle
\psi_2|A^2|\psi_1\rangle|^2,
\end{eqnarray}
where $\mid\psi_1\rangle$ and $\mid\psi_2\rangle$ are the pure-state
vectors, $A$ is an observable, and $\Delta A(\psi_1)$ is dispersion of
the observable $A$.

Our aim is to rewrite this inequality in the tomographic form. We
use the optical tomographic representation for the one-mode photon
state. The tomogram $w(X,\Theta)$ depends on the homodyne quadrature
$X$ and local oscillator phase $\Theta$; this tomogram can be
measured in the experiments with homodyne photon state detection.

If the observable $A$ in (\ref{eq.dd1.39}) is an analog of the
position operator the inequality in the tomographic form reads
\begin{eqnarray}
&&\left[\int X^2 w_1(X,\mu=1,\nu=0)\,dX\right]
\left[\int X^2 w_2(X,\mu=1,\nu=0)\,dX\right]
 \nonumber\\&&\geq\frac{1}{2\pi}\int
\tilde{w}_1(X,\mu,\nu){w}_2(-Y,\mu,\nu)\exp{(i(X+Y))}\,dX\,dY\,d\mu\,
d\nu,\label{eq.dd1.40}
\end{eqnarray}
where $\tilde{w}_1(X,\mu,\nu)$ is the symplectic tomogram of ``the
state vector" $\mid \varphi_1\rangle =A^2|\psi_1\rangle$ and $w_2(Y,\mu,\nu)$ is
symplectic tomogram of the state $|\psi_2\rangle$. If one knows wave
function $\varphi_1(y)=\langle y|\varphi_1\rangle$ and $A$ is position operator, the tomogram is
$$\tilde{w}_1(X,\mu,\nu)=\frac{1}{2\pi|\nu|}\left|\int y^2\varphi_1(y)
\exp\left(\frac{i\mu}{2\nu}y^2-\frac{iX}{\nu}y\right)\,dy\right|^2.$$

For $\mu=\cos\Theta$ and $\nu=\sin\Theta$, the symplectic tomogram
coincides with the optical tomogram. Inequality~(\ref{eq.dd1.40}) can
be expressed in terms of optical tomograms of states
$|\psi_1\rangle$ and $|\psi_2\rangle$.

\section{Probability distributions and some maps of probability vectors}
In this section we study some relations for probability distributions which are considered as probability vectors. We start
with example of classical object which can be found in four
different states $a_1$,$a_2$,$a_3$ and $a_4$ with probabilities
$p_1$,$p_2$,$p_3$ and $p_4$, respectively. The nonnegative numbers
$p_k$, $k=1,2,3,4$ satisfy the condition $\sum_{k=1}^4p_k=1$.
The numbers can be considered as the components $p_k$ of $4-$vector
$\vec{p}$ which can be called the probability vector. Also these
four numbers can be considered as coordinates of point on the plane
and the domain occupied by all the probability vectors is called
simplex. Let us consider linear maps of the probability vectors by
means of the following two $4$x$4-$matrices

\begin{eqnarray}\label{eq.1.2}
M^{(1)}=\left(
          \begin{array}{cccc}
            1 & 1 & 0 & 0 \\
            0 & 0 & 1 & 1 \\
            0 & 0 & 0 & 0 \\
            0 & 0 & 0 & 0 \\
          \end{array}
        \right);  \; \;
        M^{(2)}=\left(
          \begin{array}{cccc}
            1 & 0 & 1 & 0 \\
            0 & 1 & 0 & 1 \\
            0 & 0 & 0 & 0 \\
            0 & 0 & 0 & 0 \\
          \end{array}
        \right)
.
\end{eqnarray}

We get new probability $4-$vectors

\begin{eqnarray}\label{eq.1.3}
\vec{\wp}^{\, (1)}=M^{(1)}\vec{p}=\left(
                                  \begin{array}{c}
                                    p_1+p_2 \\
                                    p_3+p_4 \\
                                    0 \\
                                    0 \\
                                  \end{array}
                                \right); \; \;
\vec{\wp}^{\, (2)}=M^{(2)}\vec{p}=\left(
                                  \begin{array}{c}
                                    p_1+p_3 \\
                                    p_2+p_4 \\
                                    0 \\
                                    0 \\
                                  \end{array}
                                \right).
\end{eqnarray}

For these vectors the components $\wp_1^{\,(1)}$,$\wp_2^{\, (1)}$
and $\wp_1^{\,(2)}$,$\wp_2^{\, (2)}$ are the nonnegative numbers
satisfying the condition $\wp_1^{\,(1)}+\wp_2^{\, (1)}=\wp_1^{\,(2)}+\wp_2^{\, (2)}=1$.
These pairs of numbers can be considered as probability outcomes
in experiments either with two different classical coins or spin
$-1/2$ particles when one measures spin projections $m=+1/2,-1/2$ of
two spins on two different directions $\vec{n_1}$ and $\vec{n_2}$.
The analogous procedure we used to map the $4-$vectors onto the $2-$vectors and it
was considered as the method of qubit portrait of qudit states to study entanglement phenomenon of qudit states in  ~\cite{CherManko1}. All the other matrices providing the result of the map on the
probability $4-$vector with two components equal to zero can be
obtained from the matrix $M^{(1)}$ by all the permutations of rows
and columns.

It is clear that the map of vectors $\vec{p}$ realized
by permutations of the vector components provides another
$4-$vector. The map is given by the set of bistochastic
matrices $\tilde{M_s}$, $s=1,2,...,24$, where
\begin{eqnarray}\label{eq.1.5}
\tilde{M_{1}}=\left(
          \begin{array}{cccc}
            0 & 0 & 0 & 1 \\
            0 & 0 & 1 & 0 \\
            0 & 1 & 0 & 0 \\
            1 & 0 & 0 & 0 \\
          \end{array}
        \right)
\end{eqnarray}
other $23$ matrices can be obtained from $\tilde{M_{1}}$ by all the
permutations of the matrix columns and rows. There exist two specific
kinds of the linear maps of the probability vectors. One is realized
by the bistochastic matrix
\begin{eqnarray}\label{eq.1.6}
\tilde{M_{c}}=\frac{1}{4}\left(
          \begin{array}{cccc}
            1 & 1 & 1 & 1 \\
            1 & 1 & 1 & 1 \\
            1 & 1 & 1 & 1 \\
            1 & 1 & 1 & 1 \\
          \end{array}
        \right).
\end{eqnarray}
The matrix projects each probability vector $\vec{p}$ onto one
vector with all components equal to $1/4$, witch are coordinates of the simplex center. Another map
is determined by the stochastic matrix $M_{1}^{pur}$, with all rows excepting the first one containing only zero matrix elements and satisfying the equality
\begin{eqnarray}\label{eq.1.8a}
(M_{1}^{pur})^2=M_{1}^{pur}.
\end{eqnarray}
The matrix maps all the vectors $\vec{p}$ onto one vector which is
analog of a "pure state" of qudit
\begin{eqnarray}\label{eq.1.9}
M_{1}^{pur}\left( \begin{array}{c}
                p_1 \\
                p_2 \\
                p_3 \\
                p_4
              \end{array} \right)=
 \left(\begin{array}{c}
                1 \\
                0 \\
                0 \\
                0
              \end{array} \right).
\end{eqnarray}
There exist other three matrices which are obtained from the matrix
$M_{1}^{pur}$ by means of the permutations of rows. Thus the maps
$M_{k}^{pur}$ , $k=(1,2,3,4)$ project any vector $\vec{p}$ onto
vertices of the simplex. There are stochastic matrices $M^{(3)}$
which provide the maps of $4-$vectors onto probability $3-$vectors
which can be called as qutrit portrait of the state. For
example the matrix
\begin{eqnarray}\label{eq.1.10} M_{3}=\left(
          \begin{array}{cccc}
            1 & 1 & 0 & 0 \\
            0 & 0 & 1 & 0 \\
            0 & 0 & 0 & 1 \\
            0 & 0 & 0 & 0 \\
          \end{array}
        \right)
\end{eqnarray} yields the map $(p_1,p_2,p_3,p_4)\rightarrow (p_1+p_2,p_3,p_4,0)$.
All the other matrices which map one $4-$vectors $\vec{p}$ onto
probability vectors with one zero component are obtained from the
matrix $M^{(3)}$ by all permutations of rows and columns. The qubit
portrait map may be realized by other kind of stochastic matrix
which has zero elements in two rows like the matrix

\begin{eqnarray}\label{eq.1.12}
M_{4}=\left(
          \begin{array}{cccc}
            1 & 1 & 1 & 0 \\
            0 & 0 & 0 & 1 \\
            0 & 0 & 0 & 0 \\
            0 & 0 & 0 & 0 \\
          \end{array}
        \right)
\end{eqnarray} which has the property $M_{4}M_{4}=M_{1}^{pur}$.
One has the map $(p_1,p_2,p_3,p_4)\rightarrow (p_1+p_2+p_3,p_4,0,0)$.

Again other matrices of this kind providing qubit portrait of the
qudit state are obtained from $M_{4}$ by all the permutations of
rows and columns. All the stochastics matrices $M$ of linear maps of
the probability vectors $\vec{p}$ form semigroup. The stochastic
matrices providing different portraits of the qudit state form
subsemigroup of the set of all the matrices $M$. Analogous
construction of the maps of the probability vectors can be presented
for any dimension of the linear space $N$.

\section{Entropies and information}
The probability vectors can be considered as arguments of some
functions characterising the degree of randomness in the system. For
example Shanon entropy ~\cite{JRLR24} reads
\begin{eqnarray}\label{eq.1.14}
H(\vec{p})=-\sum_{k=1}^4p_k\ln p_k\equiv -\vec{p}\ln\vec{p}.
\end{eqnarray}
For the $2-$vectors (qubits) or $3-$vectors
(qutrits) one has entropies given by (\ref{eq.1.14})
where instead of $4-$vectors $\vec{p}$ one uses these $2-$vectors or
$3-$vectors. One has the following entropic inequalities. Any map $M$
which acting on the probability vector with $4$ nonzero components provides the new vectors with zero components can only
decrease entropy, i.e.

\begin{eqnarray}\label{eq.1.15}
-M\vec{p}\ln M\vec{p} \leq -\vec{p}\ln \vec{p}.
\end{eqnarray}
All the $24$ permutation matrices like $\tilde{M_{c}}$ (\ref{eq.1.6}) do not decrease
the Shannon entropy. The map given by matrix $\tilde{M}_c$  increases the entropy up to maximal value $\ln 4$.
One can consider analogous properties for qubits.
For $2-$vectors the discussed maps are given by four matrices

$M^{(1)}=\left(
           \begin{array}{cc}
             1 & 1 \\
             0 & 0 \\
           \end{array}
         \right)
$, $M^{(2)}=\left(
           \begin{array}{cc}
             0 & 0 \\
               1 & 1 \\
           \end{array}
         \right)
$, $M^{(3)}=\left(
           \begin{array}{cc}
             1 & 0 \\
             0 & 1 \\
           \end{array}
         \right)
$, $M^{(4)}=\left(
           \begin{array}{cc}
             0 & 1 \\
               1 & 0 \\
           \end{array}
         \right)
$

and the matrix $M^{(5)}=\frac{1}{2}\left(
           \begin{array}{cc}
             1 & 1 \\
               1 & 1 \\
           \end{array}
         \right)$.

First two matrices decrease the entropy up to zero. The two
permutation matrices $M^{(3)}$ and $M^{(4)}$ keep the entropy of qubit
unchanged and the bistochastic matrix $M^{(5)}$ creates maximal entropy
$\ln 2$.

One can see that among the stochastic matrices $M$ with
zeros and unit matrix elements there exists the following ordering.
Let us denote in generic case of $N-$dimensional probability vectors
such stochastic matrices with $k$ rows containing only zero matrix elements as
$M^{(N)}_k$. Then it can be easily proved that the Shannon entropies
obey inequalities
\begin{eqnarray}\label{eq.1.16}
 -\vec{p}\ln \vec{p}\geq -M_1^{(N)}\vec{p}\ln M_1^{(N)}\vec{p}\geq -M_2^{(N)}\vec{p}\ln M_2^{(N)}\vec{p}
 \geq...\geq -M_k^{(N)}\vec{p}\ln M_k^{(N)}\vec{p}\geq...\geq -M_{(N-1)}^{(N)}\vec{p}\ln
 M_{(N-1)}^{(N)}\vec{p}.
\end{eqnarray}
In particular for probability vector in qutrit case
$\vec{p}=(p_1,p_2,p_3)$ one has the inequality for nonnegative
numbers $p_k$
\begin{eqnarray}\label{eq.1.17}
-p_1\ln p_1-p_2\ln p_2-p_3\ln p_3 \geq -(p_1+p_2)\ln
(p_1+p_2)-p_3\ln p_3.
\end{eqnarray}
For any probability $N-$vector $\vec{p}$ one has

\begin{eqnarray}\label{eq.1.18}
-\sum_{k=1}^Np_k\ln p_k \geq -\sum_{k=3}^N p_k\ln p_k -(p_1+p_2)\ln
(p_1+p_2).
\end{eqnarray}
In view of permutation symmetry of the Shannon entropy the
decreasing of the entropy appears if one adds any two
$\vec{p}-$vector components.

There exists subadditivity condition
for a joint probability distribution of composite system with two
subsystems.

We present the example of such distribution for two classical 
coins. The probability $4-$vector $\vec{p}$ for such distribution
has the following indices

\begin{eqnarray}\label{eq.1.19}
p_1=p_{++},p_2=p_{+-},p_3=p_{-+},p_4=p_{--}.
\end{eqnarray}

These indices show that we have probability for two "spin projections"
to have parallel or antiparallel directions along $z-$axes. Thus we
have entropic inequality obtained by considering Shannon entropies
of separate subsystems

\begin{eqnarray}
&&[-(p_1+p_2)\ln (p_1+p_2)-(p_3+p_4)\ln (p_3+p_4)]
\nonumber\\&&+[-(p_1+p_3)\ln (p_1+p_3)-(p_2+p_4)\ln (p_2+p_4)]\geq
-p_1\ln p_1-p_2\ln p_2-p_3\ln p_3-p_4\ln p_4.\label{eq.1.20}
\end{eqnarray}

It is clear that analogous inequalities can be obtained from this
one by any permutation of four numbers $1,2,3,4$ though in this case
the sense of inequalities for entropy of the subsystems is changed.
Analogously one can get inequality

\begin{eqnarray}
&&[-p_1\ln p_1-(p_2+p_3+p_4)\ln (p_2+p_3+p_4)]+
\nonumber\\&&[-p_2\ln p_2-p_3\ln p_3-(p_1+p_4)\ln (p_1+p_4)]\geq
-p_1\ln p_1-p_2\ln p_2-p_3\ln p_3-p_4\ln p_4.\label{eq.1.21}
\end{eqnarray}

This inequality is equivalent to subadditivity condition for
$6-$dimensional probability vector $\vec{p}$ for which first two
components are zero and we consider it as probability of
qubit-qutrit system with notation for the vector $\vec{q}$ like

$q_1\equiv q_{+(1)}=0$,\, $q_2\equiv q_{+(0)}=0$,\, $q_3\equiv
q_{+(-1)}=p_1$,\, $q_4\equiv q_{-(1)}=p_2$,\, $q_5\equiv
q_{-(0)}=p_3$,\, $q_4\equiv q_{-(-1)}=p_4$.

Qubit has indices $\pm$ and qutrit has indices $+1$, \, $0$, \,
$-1$. Thus, calculating Shannon entropies for this joint probability
distribution we get inequality (\ref{eq.1.21}). It is obvious that
this inequality creates other inequalities for all the permutations
of numbers $1,2,3,4$.

For bipartite system one has the notion of mutual information which
equals to difference of left and right sides on the inequalities
(\ref{eq.1.21})

\begin{eqnarray}\label{eq.1.22}
I=p_4\ln p_4-(p_2+p_3+p_4)\ln (p_2+p_3+p_4)-(p_1+p_4)\ln (p_1+p_4).
\end{eqnarray}

This information is known to be nonnegative, i.e. $I\geq 0$. One can obtain analogs of the information which is nonnegative applying in above equality all the permutations.

\section{Qudit and qubit tomograms}
For quantum spin states or for qudits the probability vectors appear
being determined by the state density matrix $\rho$. So one has for
a qudit state the state unitary tomogram

\begin{eqnarray}\label{eq.1.23}
w(m,u)=\langle m | u\rho u^\dagger | m\rangle.
\end{eqnarray}
This tomogram was introduced in ~\cite{SudPhysLet2004}.
Here $m=-j,-j+1,...,j-1,j$, \, $j=0,1/2,1,3/2,...$ the pure state
$|m\rangle$ satisfies the eigenvalue condition

\begin{eqnarray}\label{eq.1.24}
\hat{J_z}| m\rangle = m| m\rangle
\end{eqnarray}
where $\hat{J_z}$ is spin projection on $z-$axes. The matrix $u$ is
$(2j+1)$x$(2j+1)$ unitary matrix. If the matrix $u$ is the matrix of
irreducible representation of the group $SU(2)$ the unitary tomogram
$w(m,u)$ becomes the function $w(m,\vec{n})$ where $\vec{n}$ is unit
$3-$vector determining the point on Poincare sphere $S^2$. Then the
tomogram is called spin-tomogram. The unitary and spin tommograms
satisfy the nonnegativity condition, i.e. $w(m,u)\geq0$ and
normalisation condition

\begin{eqnarray}\label{eq.1.25}
\sum_{m=-j}^j w(m,u) =\sum_{m=-j}^j w(m,\vec{n})=1.
\end{eqnarray}

The density matrix can be reconstructed if one knows the tomogram
$w(m,\vec{n})$ or $w(m,u$). In case of two qudits the unitary
tomogram of the bipartite system state determined by density matrix
$\rho(1,2)$ is defined as

\begin{eqnarray}\label{eq.1.26}
w(m_1,m_2,u)=\langle m_1m_2 | u\rho(1,2) u^\dagger | m_1m_2\rangle.
\end{eqnarray}

Here we have the spin projection $m_k$
\begin{eqnarray}\label{eq.1.27}
-j_k\leq m_k\leq j_k, \, k=1,2
\end{eqnarray}
and the matrix $u$ is $(2j_1+1)(2j_2+1)$x$(2j_1+1)(2j_2+1)$ unitary
matrix. In case of $u=u_1\bigotimes u_2$ where $u_k$ are
$(2j_k+1)$x$(2j_k+1)$ unitary matrices which are the matrices of
irreducible representations of the $SU(2)$x$SU(2)$ group the unitary
tomogram $w(m_1,m_2,u)$ becomes the spin tomogram
$w(m_1,m_2,\vec{n_1},\vec{n_2})$ where vectors $\vec{n}_k$ are unit
vectors determining the points on two Poincare spheres. If one knows
the tomograms $w(m_1,m_2,\vec{n_1},\vec{n_2})$ or $w(m_1,m_2,u)$ the
density matrix $\rho(1,2)$ can be reconstructed (see, e.g. ~\cite{IbortPS}). The
unitary and spin-tomograms can be represented as probability
vectors. For example for qudit state with density matrix $\rho$
which has the nonnegative eigenvalues
$\rho_1,\rho_2,...,\rho_{2j+1}$ and corresponding normalized
eigenvectors $\vec{u}_{01},\vec{u}_{02},...,\vec{u}_{02j+1}$ the
tomographic probability vector $\vec{w}(u)$ with components $w(m,u)$ reads
\begin{eqnarray}\label{eq.1.28}
\vec{w}(u)=|uu_0|^2\vec{\rho}.
\end{eqnarray}
Here $\vec{\rho}$ is column vector with the nonnegative components
$\rho_k$, $k=1,2,...,2j+1$. The columns of the unitary matrix $u_0$
are the vectors $\vec{u}_{ok}$. The notation $|A|^2$ for any matrix
$A$ means $|A|^2_{jk}=|A_{jk}|^2$.
Now we formulate the inequalities for the qudit state tomographic
probability vector. Applying the inequalities (\ref{eq.1.16}) to the
tomographic probability vector $\vec{w}(u)$ of qudit state we get

\begin{eqnarray}
 &&-\vec{w}(u)\ln \vec{w}(u)\geq -M_1^{(2j+1)}\vec{w}(u)\ln M_1^{(2j+1)}\vec{w}(u)\geq -M_2^{(2j+1)}\vec{w}(u)\ln M_2^{(2j+1)}\vec{w}(u)
 \nonumber\\&& \geq...\geq -M_k^{(2j+1)}\vec{w}(u)\ln M_k^{(2j+1)}\vec{w}(u)\geq...\geq -M_{(2j-1)}^{(2j+1)}\vec{w}(u)\ln
 M_{(2j-1)}^{(2j+1)}\vec{w}(u).\label{eq.1.30}
\end{eqnarray}
These inequalities take place for any unitary matrix $u$. Also for
the matrix $u$ which is the matrix of irreducible representation of
the group SU(2) the corresponding inequalities take place for any
unit vector $\vec{n}$ determining the point on Poincare sphere.
Since the minimum of the Shannon entropy corresponding to the
spin-tomographic probability vector $\vec{w}(u)$ for $u=u_0^{-1}$ is
equal to von Neuman entropy we get inequality

\begin{eqnarray}
 &&S_{VN}\geq -M_1^{(2j+1)}\vec{w}(u_0^{-1})\ln M_1^{(2j+1)}\vec{w}(u_0^{-1})\geq -M_2^{(2j+1)}\vec{w}(u_0^{-1})\ln M_2^{(2j+1)}\vec{w}(u_0^{-1})
 \nonumber\\&& \geq...\geq -M_k^{(2j+1)}\vec{w}(u_0^{-1})\ln M_k^{(2j+1)}\vec{w}(u_0^{-1})\geq...\geq -M_{(2j-1)}^{(2j+1)}\vec{w}(u_0^{-1})\ln
 M_{(2j-1)}^{(2j+1)}\vec{w}(u_0^{-1}).\label{eq.1.31}
\end{eqnarray}
Thus the von Neuman entropy provides upper bound for all the
entropies associated with the portrait tomographic probability
vectors taken in the point $u=u_0^{-1}$. It is clear that for pure
state the von Neuman entropy equals zero. Since the entropy in the
inequalities (\ref{eq.1.30}) are nonnegative it means that all these
entropies have minimal value equal to zero for $u=u_0^{-1}$. Let us
consider expression for information $I$ (\ref{eq.1.22}) where we
interpret probability vector as tomogram of the qudit state
corresponding to $j=3/2$. Then the nonnegativity of the information
$I\geq$ $0$ gives inequality
\begin{eqnarray}
 && w(-\frac{3}{2}, u)\ln w(-\frac{3}{2}, u)-[ w(\frac{1}{2}, u)+ w(-\frac{1}{2},
 u)+ w(-\frac{3}{2}, u)]\ln[ w(\frac{1}{2}, u)+w(-\frac{1}{2}, u)w(-\frac{3}{2},
 u)]
 \nonumber\\&&-[w(\frac{3}{2}, u)+w(-\frac{3}{2}, u)]\ln[w(\frac{3}{2}, u)+w(-\frac{3}{2},
 u)]\geq 0.
 \label{eq.1.32}
\end{eqnarray}
It means that for the point $u_0^{-1}$ we have condition of
positivity of the "tomographic information" provided the von Neuman
entropy is given. The physical meaning of this inequality needs
extra clarification. If the probability vector in (\ref{eq.1.22})
corresponds to two-qubit state the information nonnegativity gives
\begin{eqnarray}
 && w(-\frac{1}{2}, -\frac{1}{2},u)\ln w(-\frac{1}{2}, -\frac{1}{2},)-[ w(\frac{1}{2}, -\frac{1}{2}, u)+ w(-\frac{1}{2},\frac{1}{2},
 u)+ w(-\frac{1}{2},-\frac{1}{2}, u)]
 \nonumber\\&&\ln[w(\frac{1}{2}, -\frac{1}{2}, u)+ w(-\frac{1}{2},\frac{1}{2},
 u)+ w(-\frac{1}{2},-\frac{1}{2}, u)]
 \nonumber\\&&-[w(\frac{1}{2}, \frac{1}{2},  u)+w(-\frac{1}{2},-\frac{1}{2}, u)]\ln[w(\frac{1}{2}, \frac{1}{2},  u)+w(-\frac{1}{2},-\frac{1}{2}, u)]\geq 0.
 \label{eq.1.33}
\end{eqnarray}
For pure two qubit state which is entangled state violating Bell
inequality ~\cite{Bell64} one can consider specific inequality for the
"information" which provides some relation for the given probability
$w(m_1,m_2,\vec{n}_1,\vec{n}_2)$. It means that there exists some
correlation of the violation of Bell inequality for particular
unitary matrix $u=u_1\bigotimes u_2$ corresponding to directions
$\vec{n}_1,\vec{n}_2$ and the information inequality.

\section{Tomographic cumulants}
Any probability distribution is characterized by specific 
numbers, like Shannon entropy, moments of random variables, etc. One
of such characteristics is cumulant. For given probability
distribution $W(X)$ of continuous variable $X$ the cumulants are
defined as
\begin{eqnarray}\label{eq.1.34}
g(t)=\ln \langle\exp(tX)\rangle=\sum_{n=2}^nK_n\frac{t^n}{n!}.
\end{eqnarray}
Here
\begin{eqnarray}\label{eq.1.35}
 \langle\exp(tX)\rangle=\int W(X)\exp(tX)dx
\end{eqnarray}
and cumulants $K_n$ are coefficients in the series. In probability
representation of quantum mechanics the states are described by
symplectic tomogram $M(X,\mu,\nu)$ which is probability distribution
of homodyne quadrature $X$ depending on two real parameters $\mu$
and $\nu$. Thus we introduce the tomographic cumulants
$K_n(\mu,\nu)$ which are given by the formula
\begin{eqnarray}\label{eq.1.36}
g(t,\mu,\nu)=\ln\int M(X,\mu,\nu)\exp(tX)dX=\sum_{n=1}^\infty
t^n\frac{K_n(\mu,\nu)}{n!}.
\end{eqnarray}
For optical tomogram $w(X,\Theta)=M(X,\cos\Theta,\sin\Theta)$ the
cumulants are defined by generating function
\begin{eqnarray}\label{eq.1.37}
g(t,\Theta)=\ln\int w(X,\Theta)\exp(tX)dX=\sum_{n=1}^\infty
t^n\frac{K_n(\Theta)}{n!}.
\end{eqnarray}
Let us introduce the function $C(t,\Theta)$ which we will use as a
characteristic of state gaussianity
\begin{eqnarray}\label{eq.1.38}
C(t,\Theta)=\ln\int w(X,\Theta)\exp(tX)dX-t\int
Xw(X,\Theta)dX-\frac{t^2}{2}[\int X^2w(X,\Theta)dX-(\int
Xw(X,\Theta)dX)^2].
\end{eqnarray}
In experiments with homodyne photon detection the optical tomogram
$w(X,\Theta)$ is measured. For gaussian photon states the introduced
function must be equal to zero. The deviation of this function of
parameter $t$ and local oscillator phase $\Theta$ from zero gives
the information on nongaussianity degree of the quantum state. This
characteristics can be easily extracted from the experimental
homodyne detection data. One can introduce the parameter
of nongaussianity
$${\rm Ch}=\int_0^\infty\int_0^{2\pi} C(t,\Theta)e^{-t}\,dt\,d\Theta.$$
For the Gaussian state, it is equal to zero.
\section{Conclusion}
To resume we point out the main results of our work. We expressed the state-extended uncertainty relations for two states of the photon in tomographic form providing some inequalities which can be checked experimentally. Some new entropic inequalities for spin-tomograms are obtained including inequality for bipartite system and inequality for particle with spin equal $3/2$.

We introduced tomographic cumulant as parameter which can be measured in experiments on homodyne photon detection ~\cite{Bellini1} and it provides the characteristics of degree of nongaussianity of the photon state.

\section*{Acknowledgements}
VIM thanks Organizers of conference for the support.
This study was partially supported by the Russian Foundation for
Basic Research under Projects Nos.~10-02-00312 and 11-02-00456. The
travel Grant~No.12-02-09251 of the Russian Foundation for Basic Research is
acknowledged.

\end{document}